\documentclass[a4paper,12pt]{article}

\usepackage{lscape}		

\usepackage{cmap}						
\usepackage[T2A]{fontenc}				
\usepackage[utf8]{inputenc}				
\usepackage[english, russian]{babel}	

\usepackage{amsthm,amsfonts,amsmath,amssymb,amscd} 

\usepackage{indentfirst} 

\usepackage[usenames]{color}
\usepackage{color}
\usepackage{colortbl}

\usepackage{longtable}					
\usepackage{multirow,makecell,array}	

\usepackage[singlelinecheck=off,center]{caption}	
\usepackage{soul}									

\usepackage{cite} 

\usepackage[plainpages=false,pdfpagelabels=false]{hyperref}
\definecolor{linkcolor}{rgb}{0.9,0,0}
\definecolor{citecolor}{rgb}{0,0.6,0}
\definecolor{urlcolor}{rgb}{0,0,1}
\hypersetup{
    colorlinks, linkcolor={linkcolor},
    citecolor={citecolor}, urlcolor={urlcolor}
}

\usepackage{graphicx}		
\graphicspath{{pictures/}}	

\usepackage{float}

\makeatletter
\renewcommand{\@biblabel}[1]{#1.}	
\makeatother

\usepackage{footnote}

\providecommand{\keywords}[1]{\textbf{\textit{Ключевые слова:}} #1}

\title{Математическое моделирование и прогнозирование COVID-19 в Москве и Новосибирской области\footnote{Работа выполнена при поддержке Российского научного фонда (проект №~18-71-10044), а именно постановка, анализ идентифицируемости и численное решение задачи прогнозирования для математической модели SEIR-HCD (разделы 1 и 3), а также при поддержке Математического Центра в Академгородке в рамках соглашения № 075-15-2019-1675 с Министерством науки и высшего образования Российской Федерации, а именно постановка и численное решение обратной задачи для математической модели SEIR-D (разделы 2 и 3).}}

\author{Криворотько О.И., Кабанихин С.И., Зятьков Н.Ю., \\ Приходько А.Ю., Прохошин~Н.М., Шишленин М.А.\\
\small{Институт вычислительной математики и математической геофизики СО РАН},\\
\small{Математический центр в Академгородке, Россия}, \\
\small{Новосибирский государственный университет, Новосибирск, Россия}}
\date{}

\begin{document}
\maketitle

\begin{abstract}
В работе сформулированы и решены задачи уточнения неизвестных параметров математических моделей распространения коронавирусной инфекции COVID-19, основанных на моделях SEIR типа, по дополнительной информации о количестве выявленных случаев, смертности, коэффициента самоизоляции и проведенных тестов для города Москвы и Новосибирской области с 23.03.2020. В рамках используемых моделей популяция разделена на семь (SEIR-HCD) и пять (SEIR-D) групп со схожими признаками с вероятностями перехода между группами, зависящими от конкретного региона. Проведен анализ идентифицируемости математической модели SEIR-HCD, который выявил наименее чувствительные неизвестные параметры к дополнительным измерениям. Задачи уточнения параметров сведены к задачам минимизации соответствующих целевых функционалов, которые решены с помощью стохастических методов (имитации отжига, дифференциальной эволюции, генетического алгоритма и др.). Для разного количества тестируемых данных разработан прогностический сценарий развития заболевания в городе Москве и Новосибирской области, предсказан пик развития эпидемии в Москве с погрешностью в 2 дня и 174 выявленных случая и проведен анализ применимости разработанных моделей.
\end{abstract}


\section*{Введение}
\begin{figure}[H]
  \center{\includegraphics[width=0.8\linewidth]{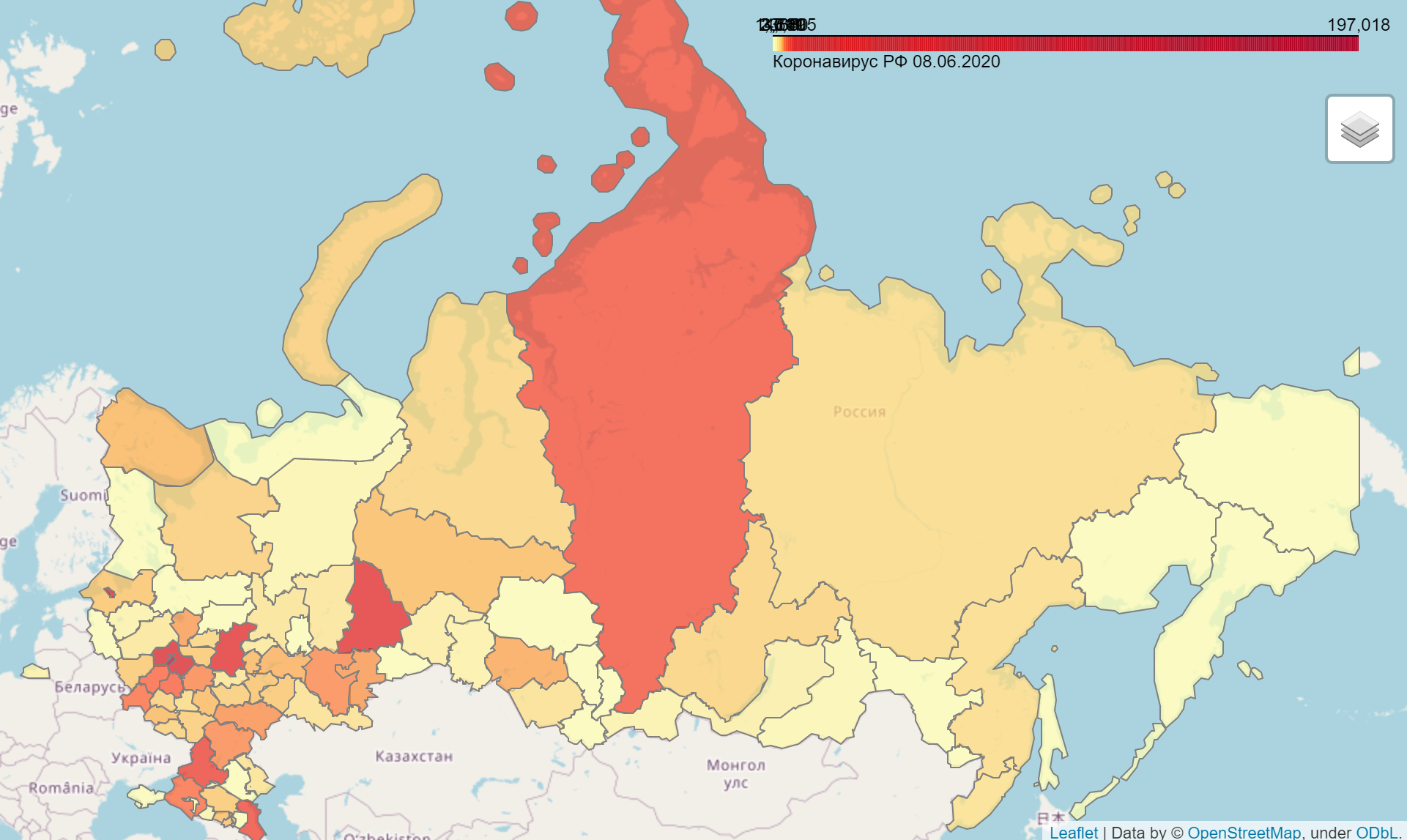}}
  \caption{Количество выявленных случаев в регионах Российской Федерации на 08.06.2020.} 
  \label{ris:Russia_map_08-06-2020}
\end{figure}

В декабре 2019 года произошла вспышка пневмонии в Ухане 2019-2020 годов, в результате которой был впервые обнаружен штамм COVID-19 при анализе нуклеиновой кислоты у пациента с пневмонией. К первой декаде июня 2020 года пандемия охватила 188 стран, в которых было выявлено более 7 миллионов случаев заражения, 411000 человек из которых погибли. Российская Федерация по числу выявленных случаев находится на 3 месте в рейтинге стран после США и Бразилии с 484630 случаями заражения на 9 июня 2020 года (см. Таблицу~\ref{TAB:top5_countries_COVID}). Несмотря на пройденный пик выявления заболеваемости в мире уменьшение количества выявленных случаев не наблюдается в течение достаточно длительного периода~\cite{Hopkins_data}. На рис.~\ref{ris:Russia_map_08-06-2020} приведена карта распределения выявленных случаев в регионах Российской Федерации на 08.06.2020. Разработка сценариев развития заболевания в регионах г. Москва (наибольшее число выявленных случаев) и Новосибирская область (средний показатель выявления) является важным шагом для принятия соответствующих мер по сдерживанию эпидемии в регионах.
\begin{savenotes}
    \begin{table}[!ht]
    \caption{Статистика по распространенности коронавируса COVID-19 для пяти стран-лидеров по числу выявленных случаев на 9 июня 2020 года.}
    \label{TAB:top5_countries_COVID}
        \begin{tabular}{lp{2.3cm}p{3cm}p{2cm}p{2.5cm}}
            \hline\noalign{\smallskip}
            Страна & \multicolumn{2}{c}{Количество зараженных случаев} & Количество смертей & Количество вылеченных\\
            {} & Всего & Новые случаи & {} & {} \\
            \noalign{\smallskip}\hline\noalign{\smallskip}
            США & 1~961~185 & 17~250 & 111~007 & 524~855 \\
            Бразилия & 707~412 & 15~654 & 37~134 & 396~737 \\
            Россия & 485~253 & 8~595 & 6~141 & 241~917 \\
            Великобритания & 288~834 & 1~213 & 40~680 & 1~257 \\
            Индия & 267~046 & 8~442 & 7~473 & 134~165 \\
            \noalign{\smallskip}\hline
        \end{tabular}
    \end{table}
\end{savenotes}

В работах~\cite{Tamm_2020, Koltsova_2020} впервые описаны сценарии развития эпидемии коронавируса в городе Москва и показано, что при вводимых ограничениях Правительством РФ предполагается растягивание эпидемии более чем на год, а в случае более жестких мер удастся подавить эпидемию и существенно снизить число умерших, однако в популяции не будет вырабатываться групповой иммунитет, и популяция остается уязвимой для повторных вспышек эпидемии.

Описание распространения эпидемии коронавируса COVID-19 в Китае с учетом пассажиропотоков, влияния вмешательства здравоохранения и инкубационного периода в рамках математических моделей, основанных на SEIR структуре, проводилось группами китайских ученых~\cite{Zlojutro_2019, Cheng_JIIP2020, Tang_2020}.

Цель данной работы --- исследовать зависимость достоверности прогнозов SEIR-моделей от количества и качества информации о распространении пандемии COVID-19.

Отметим, что группа ученых Сингапурского университета технологии и дизайна рассчитала, что эпидемия коронавируса в России закончится 17 августа 2020 года. Они опубликовали результаты своих расчетов, но так и не обнародовали детали, упомянув, что используют SIR-модель.

Модель SIR (Susceptible, Infected, Recovered) является базовой для описания распространения инфекционных заболеваний и была предложена в 1920-х годах шотландскими эпидемиологами Андерсоном Кермаком и Уильямом Маккендриком. Согласно SIR, население делится на три группы: восприимчивые ($S$), инфицированные (зараженные) ($I$) и выздоровевшие ($R$). С течением времени возможны переходы $S \rightarrow I$ (заражение) и $I \rightarrow R$ (выздоровление или смерть). SIR-модель перестает работать, если необходимо учитывать больше данных. Например, различную плотность населения в разных районах или разные пути передачи инфекции. Из-за очевидных недостатков SIR многократно дорабатывалась. Сегодня существует целое семейство моделей (и даже открытых кодов для расчетов по ним), разработанных на базе SIR-моделей~\cite{Noll_2020}:
\begin{itemize}
\itemsep0em 
    \item SIRS: <<Восприимчивые $\rightarrow$ инфицированные $\rightarrow$ выздоровевшие $\rightarrow$ восприимчивые>>. Модель для описания динамики заболеваний c временным иммунитетом.
    \item SEIR: <<Восприимчивые $\rightarrow$ контактировавшие с инфекцией (Exposed) $\rightarrow$ инфицированные $\rightarrow$ выздоровевшие>>. Модель для описания распространения заболеваний с инкубационным периодом.
    \item SIS: <<Восприимчивые $\rightarrow$ инфицированные $\rightarrow$ восприимчивые>>. Модель для распространения заболевания, к которому не вырабатывается иммунитет.
    \item MSEIR: <<Наделенные иммунитетом от рождения (Maternally derived immunity) $\rightarrow$ восприимчивые $\rightarrow$ контактировавшие с инфекцией (Exposed) $\rightarrow$ инфицированные $\rightarrow$ выздоровевшие>>. Модель, учитывающая иммунитет детей, приобретенный внутриутробно.
\end{itemize}

В работе используются две модели: SEIR-HCD (которую используют также во Франции и в Германии) и SEIR-D. В SEIR-HCD популяция делится на семь групп. К четырем традиционным SEIR, добавляются $H$ -- госпитализированные, $C$ -- критические (подключенные к аппарату вентиляции легких), $D$ -- умершие. В SEIR-D популяция делится на пять групп. К четырем традиционным SEIR добавляется $D$. Выбор вариации моделей на базе SEIR не случайный, так как моделируемое заболеваний COVID-19 обладает достаточно длительным инкубационным периодом (5-14 дней), в период которого носители не проявляют симптомов заболевания, но являются инфицированными~\cite{WHO_2020}.

В работе сначала уточняются коэффициенты перехода из одной группы в другую посредством решения обратной задачи с использованием данных о количестве протестированных заразившихся $I(t)$ и $E(t)$, а также умерших $D(t)$. После этого мы решаем прямую задачу~-- рассчитываем сценарий развития эпидемии.

Мы также используем информацию о заболевании:

- длительность инкубационного периода,

- продолжительность латентного периода (время с момента инфицирования до того момента, когда инфицированный становится распространителем инфекции),

- параметр контагиозности и другие.

В результате обобщения модели мы можем рассчитать сценарий развития заболевания в конкретных городах России. Например, сказать, когда ожидается пик заболеваемости в Москве и Новосибирской области, когда удастся выйти на плато, и каким ожидается спад эпидемии в каждом конкретном случае. В математических моделях не учитываются:

- социальное дистанцирование,

- климатические условия,

- возрастные характеристики,

- особенности иммунитета человека,

- пассажиропотоки между городами РФ и другие.

В работе приведено математическое описание SEIR-HCD и SEIR-D моделей, приведены постановки обратных задач (уточнения параметров для г. Москвы и Новосибирской области) и алгоритмы их решения. В заключении приведены качественное и количественное сравнение двух математических моделей, условия их использования и анализ полученных прогнозов развития эпидемии COVID-19 в г.~Москва и Новосибирской области.

\section[Математическая модель SEIR-HCD]{Математическая модель SEIR-HCD}

\begin{figure}[H]
  \begin{minipage}[h]{0.49\linewidth}
  \center{\includegraphics[width=0.8\linewidth]{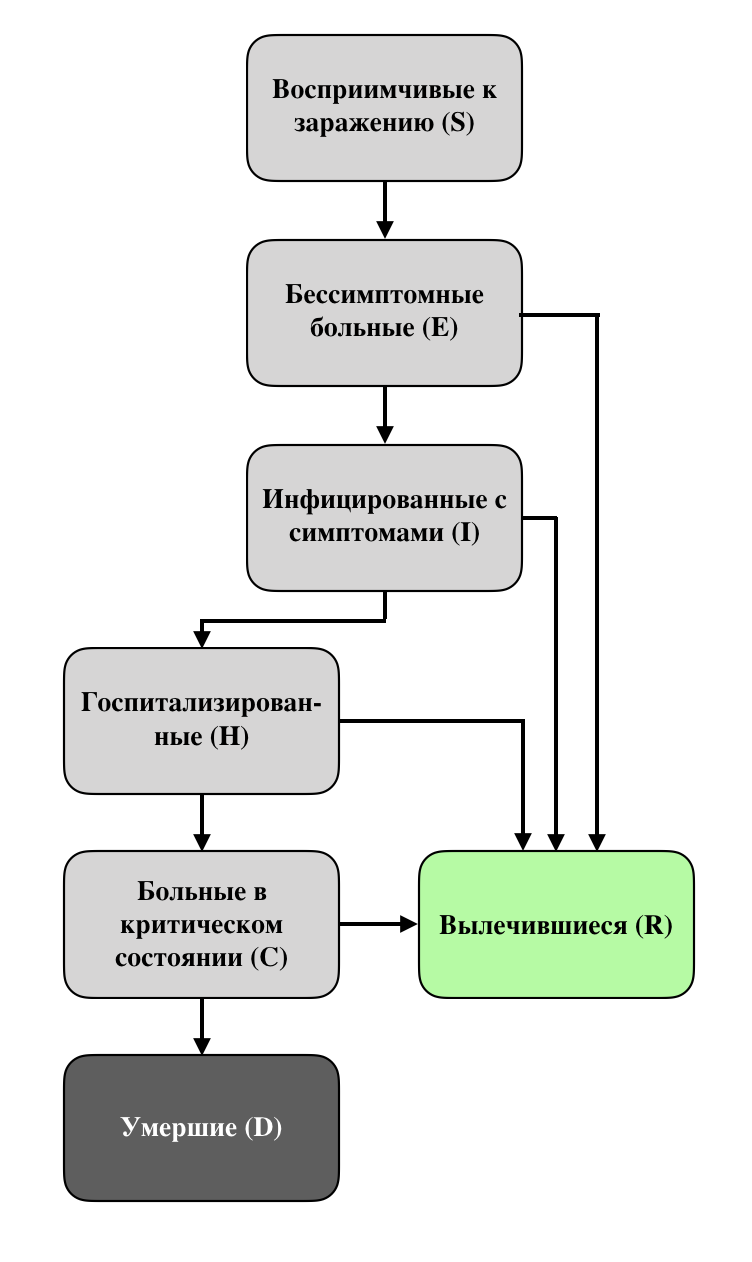}}
  \end{minipage}
  \begin{minipage}[h]{0.49\linewidth}
  \center{\includegraphics[width=0.8\linewidth]{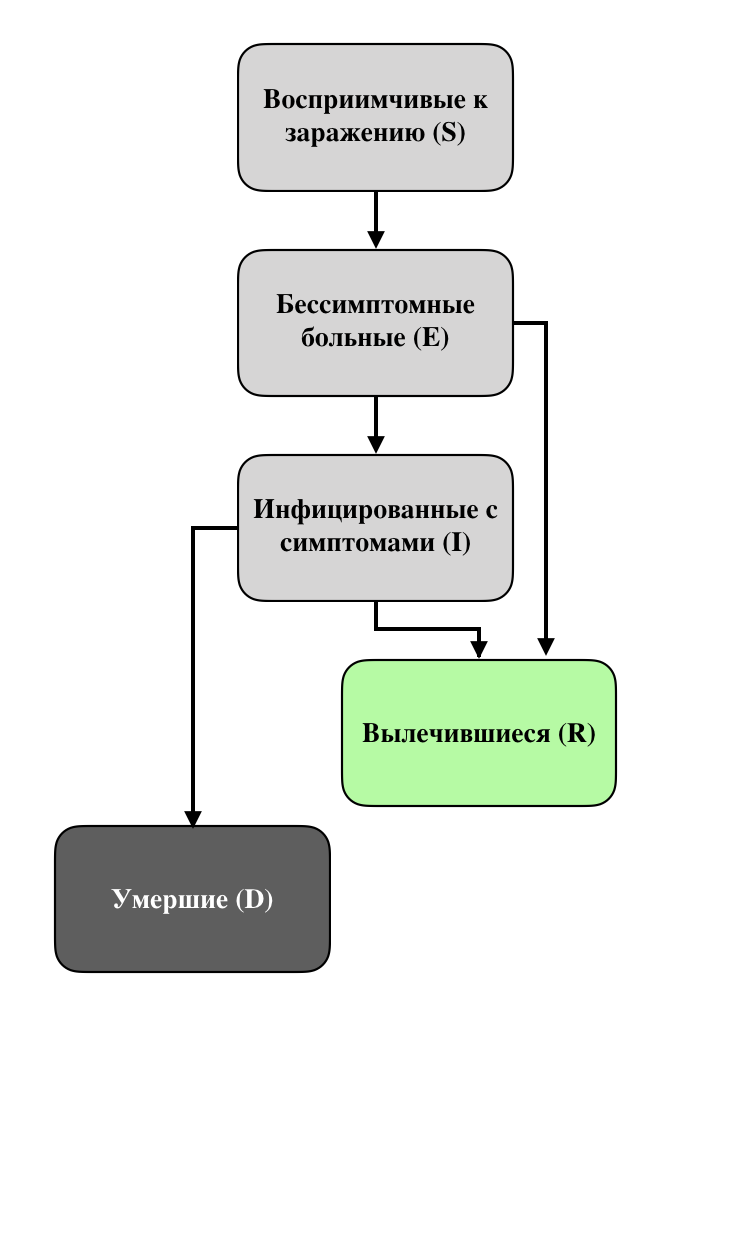}}
  \end{minipage}
  \caption{Схемы математических моделей SEIR-HCD~(\ref{model_COVID}) слева и SEIR-D~(\ref{model_2}) справа.} 
  \label{ris:SEIR-HCD_model_scheme}
\end{figure}

Математическая модель SEIR-HCD распространения коронавируса COVID-19 в конкретном регионе Российской Федерации впервые предложена в работе~\cite{SEIRHCD_model2020} и основана на системе из 7 нелинейных обыкновенных дифференциальных уравнений на отрезке $t\in [t_0,T]$:
\begin{eqnarray}\label{model_COVID}
\left\{\begin{array}{ll}
\dfrac{dS}{dt} = - \dfrac{5-a(t-\tau)}{5}\left(\dfrac{\textcolor{red}{\alpha_I} S(t)I(t)}{N(t)} + \dfrac{\textcolor{red}{\alpha_E} S(t)E(t)}{N(t)}\right) + \gamma R(t),\\[7pt]
\dfrac{dE}{dt} = \dfrac{5-a(t-\tau)}{5}\left(\dfrac{\textcolor{red}{\alpha_I} S(t)I(t)}{N(t)} + \dfrac{\textcolor{red}{\alpha_E} S(t)E(t)}{N(t)}\right) - (\textcolor{red}{\kappa} +\rho) E(t),\\[7pt]
\dfrac{dI}{dt} = \textcolor{red}{\kappa} E(t) - (\textcolor{red}{\beta}+\textcolor{red}{\nu})I(t),\\[7pt]
\dfrac{dR}{dt} = \textcolor{red}{\beta} I(t) + \rho E(t) - \gamma R(t) + \varepsilon_{HR}H(t),\\[7pt]
\dfrac{dH}{dt} = \textcolor{red}{\nu} I(t) + \textcolor{red}{\varepsilon_{CH}} C(t) - (\varepsilon_{HC} + \varepsilon_{HR}) H(t),\\[7pt]
\dfrac{dC}{dt} = \varepsilon_{HC} H(t) - (\textcolor{red}{\varepsilon_{CH}}+\textcolor{red}{\mu})C(t),\\[7pt]
\dfrac{dD}{dt} = \textcolor{red}{\mu} C(t);\\[7pt]
\end{array}\right.
\end{eqnarray}
с начальными условиями
\begin{eqnarray}\label{init_data}
\begin{array}{cc}
    S(t_0)=S_0,\, E(t_0)=\textcolor{red}{E_0},\, I(t_0)=I_0,\, R(t_0)=R_0,\\
    H(t_0) = H_0,\, C(t_0)=C_0,\, D(t_0) = D_{0}.
\end{array}
\end{eqnarray}
Вся популяция $N = S+E+I+R+H+C+D$ состоит из следующих 7 групп (схема модели представлена на рис.~\ref{ris:SEIR-HCD_model_scheme} слева):
\begin{itemize}
\itemsep0em 
    \item $S$~-- восприимчивые (незараженные) индивидуумы c 3 лет;
    \item $E$~-- зараженные индивидуумы или находящиеся в инкубационном периоде;
    \item $I$~-- инфицированные индивидуумы с симптомами;
    \item $R$~-- вылеченные индивидуумы;
    \item $H$~-- госпитализированные, т.е. с тяжелым протеканием болезни;
    \item $C$~-- находящиеся в критическом состоянии, требующие подключения аппарата ИВЛ;
    \item $D$~-- летальные случаи заболевания среди населения.
\end{itemize}
Описание параметров и их усредненные значения приведены в Таблице~\ref{TAB:parameter_info}.

\subsection{Постановка обратной задачи}
Предположим, что известны дополнительные измерения о 3 функциях в фиксированные моменты времени:
\begin{eqnarray}\label{IP_data}
    E(t_k) = 0.42 f_k,\, I(t_k) = 0.58f_{k},\, \Delta D(t_k) = g_{k},\quad t_k\in(t_0,T),\, k=1,\ldots, K,
\end{eqnarray}
где $f_k$~-- количество выявленных больных в день $k$, $g_k$~-- количество умерших в результате заболевания в день $k$, $\Delta D(t_k) = D(t_{k}) - D(t_{k-1})$, $K$~-- количество дней в статистике. В модели предполагаем, что около 42\% выявленных не проявляли симптомов.

Неизвестные параметры: \textcolor{red}{$q=(\alpha_E, \alpha_I, \kappa, \beta, \nu, \varepsilon_{CH}, \mu, E_0)$} $\in\mathbb{R}^{8}$.

\underline{Обратная задача}~(\ref{model_COVID})-(\ref{IP_data}) состоит в определении вектора параметров $q$ по дополнительным измерениям~(\ref{IP_data}).

Обратная задача сводится к задаче минимизации целевого функционала:
\begin{eqnarray}\label{func_SEIR-HCD}
    J(q) = \sum\limits_{k=1}^K \Big[ (c^{test}_k E(t_k;q) - 0.42f_k)^2 +  (c^{test}_k I(t_k;q)-0.58f_k)^2 + ( \Delta D(t_k;q)-g_k)^2 \Big].
\end{eqnarray}
Здесь $c^{test}_k\in [0,1]$~-- отношение количества сделанных тестов к размеру здоровой популяции региона в день $k$. Функционал~(\ref{func_SEIR-HCD}) построен на следующем соображении: в среднем около 42\% выявленных случаев не проявляют симптомов (группа $E(t)$), в результате чего общее число выявленных случаев пропорционально 42\% протестированных бессимптомных и 58\% протестированных симптомных с коэффициентом тестирования $c_k^{test}$. Так как индивидуумы, перешедшие в группу $D(t)$, остаются там с течением времени, то изменение в данный группе вычисляется как $D(t_{k}) - D(t_{k-1})$.

\subsection{Анализ чувствительности SEIR-HCD}\label{sec_identifiability}
Используя программный пакет DAISY~\cite{DAISY_identifiability} по анализу структурной идентифицируемости модели~(\ref{model_COVID})-(\ref{init_data}), показано, что математическая модель распространения коронавируса в популяции является идентифицируемой. Однако, необходимо установить чувствительность параметров $q$ к функциям модели с целью контроля устойчивости полученного решения обратной задачи и качества прогнозирования~\cite{ADV_KOI_2020}.

В качестве первого шага мы попытаемся оценить те параметры, к которым решение модели наиболее чувствительно. Они, в свою очередь, определяются с помощью расчета полу относительной чувствительности. 

Чтобы описать эту методологию, предположим, что мы хотим определить относительную чувствительность наблюдаемых модельных величин $x = (E,I,D)^T$ (и, следовательно, модельного решения $\bar{x}=(S,E,I,R,H,C,D)^T$) к конкретным параметрам $q_k, k = 1, \ldots, 8$. Полуотносительная чувствительность модельного решения к параметру $q_k$ определяется выражением $\frac{\partial x_i(t;q)}{\partial q_k} \cdot q_k$ и вычисляется путем формального дифференцирования модели ОДУ:
\begin{eqnarray}\label{ODE}
\left\{\begin{array}{ll}
\dfrac{d\bar{x}}{dt} = f(t,\bar{x};q),\\
\bar{x}(0) = \bar{x}_0
\end{array}\right.
\end{eqnarray}
по $q_k$ с изменением порядка дифференцирования по времени и параметрам.

Таким образом, мы получаем $(7 \times 8)$-мерную систему дифференциальных уравнений для функции чувствительности $\bar{x}_q(t;q)=\partial \bar{x}(t;q)/\partial q$:
\begin{eqnarray*}
\dfrac{d}{dt}\left(\dfrac{\partial \bar{x}(t)}{\partial q}\right) =\dfrac{\partial f}{\partial \bar{x}} \dfrac{\partial \bar{x}(t)}{\partial q} + \dfrac{\partial f}{\partial q},
\end{eqnarray*}
с начальными условиями:
\begin{eqnarray*}
\dfrac{\partial \bar{x}(0)}{\partial q} = \dfrac{\partial \bar{x}_0}{\partial q} = 0.
\end{eqnarray*}
Последнее означает, что нулевая матрица при любом начальном условии не является функцией всех оцененных параметров. Здесь $\partial f /\partial \bar{x}$ -- якобиан системы ОДУ, а $\partial f / \partial q$ -- производная правой части по рассматриваемым параметрам.

\begin{small}
   \begin{table}[h!]
	\begin{center}
		\caption{Полуотносительные чувствительности различных состояний модели к параметрам, отсортированные по убыванию.}
	\begin{tabular}{ p{2.75 cm}  p{2.35 cm}  p{2.3 cm} p{2.75 cm}  p{2.35 cm}  p{2.3 cm}}
\hline
 Переменная $x_i$ & Параметр $q_k$ & $\|c_i(t)\frac{\partial x_i(t)}{\partial q_k}q_k\|_2$ & Переменная $x_i$ & Параметр $q_k$ & $\|c_i(t)\frac{\partial x_i(t)}{\partial q_k}q_k\|_2$\\[5pt]
 \hline
	 $D$& $\alpha_E$ & $225.81$ & $D$& $\beta$ & $13.60$\\[5pt]
	 $I+E$& $\alpha_E$ & $156.69$ & $D$& $\varepsilon_{CH}$ & $12.64$\\[5pt]
	 $D$& $\mu$ & $89.06$&  $I+E$& $\nu$ & $3.38$\\[5pt]
	 $D$& $\rho$ & $63.27$&$I+E$& $\beta$ & $1.21$\\[5pt]
	 $I+E$& $\rho$ & $57.73$& $D$&$\alpha_I$ & $0.22$\\[5pt]
	 $D$& $\nu$ & $53.78$&$I+E$& $\alpha_I$ & $0.19$\\[5pt]
	 $D$& $\kappa$ & $47.44$&$I+E$& $\mu$ & $0.00$\\[5pt]
	 $I+E$& $\kappa$ & $30.42$&$I+E$& $\varepsilon_{CH}$ & $0.00$\\[5pt]
	\hline
	\end{tabular}
	  \label{sensitivity}
	\end{center}
	\end{table}
\end{small}

Этот процесс дает информацию о чувствительности как функцию времени на интересующем интервале. Мы хотим иметь некоторую общую меру чувствительности решения к параметрам, поэтому для каждой комбинации состояние/параметр мы берем норму (в пространстве $L_2$) по времени, а затем ранжируем полученные скаляры, чтобы определить наиболее чувствительные параметры (см. Таблицу~\ref{sensitivity}). Чем меньше значение $\|\frac{\partial x_i(t)}{\partial q_k}\|_2$, тем меньше влияние параметра $q_k$ на переменную $x_i$. Например, качество определения параметров $\varepsilon_{CH}$ и $\mu$ при решении обратной задачи не зависит от имеющихся измерений количества инфицированных $I(t)+E(t)$ в отличие, например, от более чувствительных к этим данным коэффициентов $\alpha_E, \rho, \kappa$.

\begin{figure}[H]
  \begin{minipage}[h]{0.49\linewidth}
  \center{\includegraphics[width=1\linewidth]{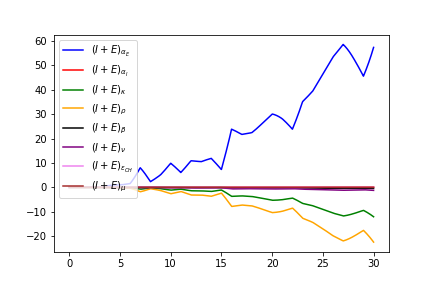}}\\ а) $x_i(t) = I(t) + E(t)$.
  \end{minipage}
  \hfill
  \begin{minipage}[h]{0.49\linewidth}
  \center{\includegraphics[width=1\linewidth]{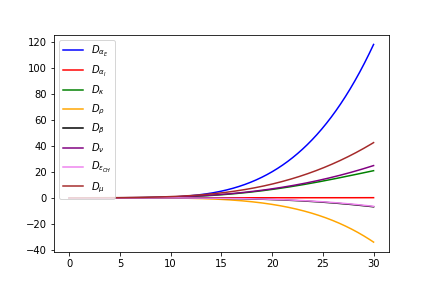}}\\ б) $x_i(t) = D(t)$.
  \end{minipage}
  \vfill
  \caption{Функция чувствительности $c^{test}(t)\frac{\partial x_i(t)}{\partial q_k}q_k$ для временного периода c 23.03.2020 по 21.04.2020 (по горизонтали 30 дней).} 
  \label{ris:Sens_30}
\end{figure}

На рисунках~\ref{ris:Sens_30} представлены графики изменения от времени функции чувствительности $c^{test}(t)\frac{\partial x_i(t)}{\partial q_k}q_k$ в зависимости от варьируемого параметра. Чем более изменчив параметр в динамике, тем чувствительность к данным измерениям выше, а, значит, определяться он будет более устойчиво. В обратной задаче~(\ref{model_COVID})-(\ref{IP_data}) самыми идентифицируемыми параметрами оказались $\alpha_E$, $\rho$ и $\kappa$, а наименее чувствительными к данным~-- параметры $\mu$, $\varepsilon_{CH}$ и $\alpha_I$.

\begin{figure}[H]
\center{\includegraphics[width=0.8\linewidth]{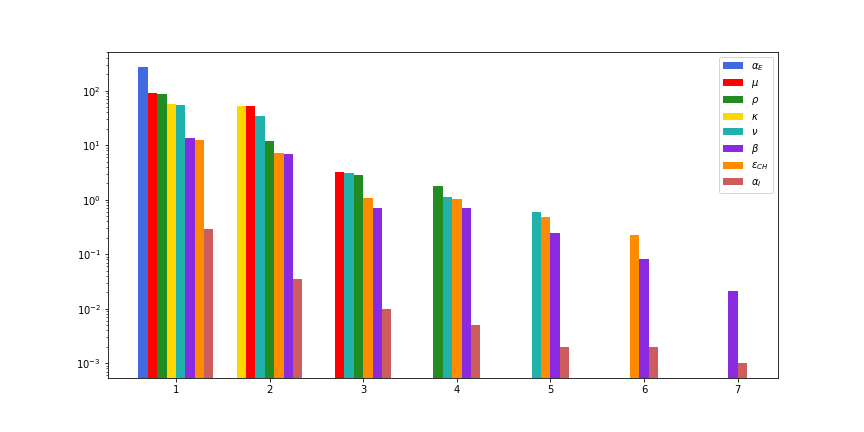} }
\caption{Величины норм перпендикуляров для каждого параметра на различных итерациях ортогонального алгоритма~\cite{ADV_KOI_2020} для математической модели~(\ref{model_COVID}).}
\label{hist_orth1_SEIR-HCD}
\end{figure}

\begin{figure}[H]
\begin{minipage}[h]{0.35\linewidth}
\center{\includegraphics[width=1\linewidth]{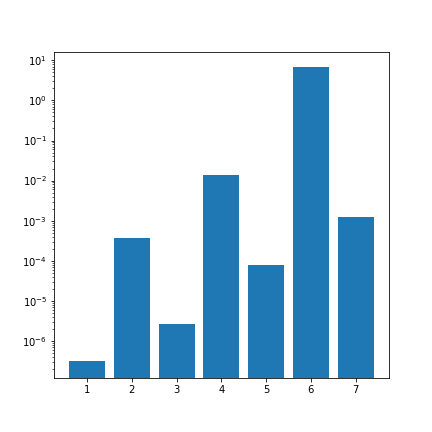} \\ а)} 
\end{minipage}
\hfill
\begin{minipage}[h!]{0.7\linewidth}
\center{\includegraphics[width=1\linewidth]{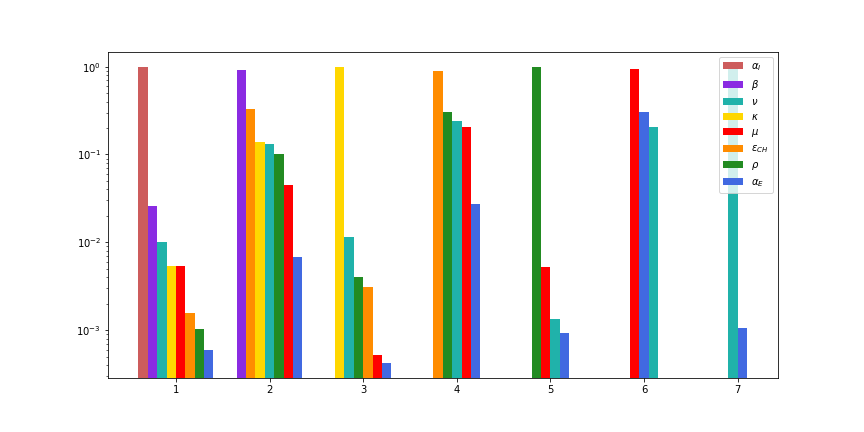} \\ б)}
\end{minipage}
\caption{а) Значение минимального собственного числа матрицы Гессиана на каждой итерации метода собственных значений б) Абсолютные значения элементов собственного вектора, соответствующего минимальному собственному значению, полученные на каждой итерации  метода собственных значений для математической модели~(\ref{model_COVID}).}
\label{Min_eigen_values_SEIR-HCD}
\end{figure}

На рисунках~\ref{hist_orth1_SEIR-HCD}-\ref{Min_eigen_values_SEIR-HCD} приведены результаты анализа чувствительности параметров математической модели~(\ref{model_COVID}) при различных итерациях ортогонального алгоритма и метода собственных значений, соответственно~\cite{ADV_KOI_2020}. Показано, что наиболее идентифицируемыми также оказались параметры заражения между бессимптомной и восприимчивой группами населения $\alpha_E$, смертности $\mu$ и скорости восстановления выявленных случаев $\rho$.

\begin{small}
   \begin{table}[h!]
	\begin{center}
		\caption{Последовательности параметров, полученные с помощью методов анализа чувствительности для математической модели~(\ref{model_COVID}), расположенные от наиболее до наименее чувствительного параметра.}
	\begin{tabular}{ p{4 cm} p{5 cm} p{5.5 cm}}
\hline
 & Ортогональный метод & Метод собственных значений \\[5pt]
 \hline
	Последовательность параметров & $\alpha_I$, $\beta$, $\varepsilon_{CH}$, $\nu$, $\rho$, $\mu$, $\kappa$, $\alpha_E$ &
	$\alpha_I$, $\beta$, $\kappa$, $\varepsilon_{CH}$, $\rho$, $\mu$, $\nu$, $\alpha_E$\\[5pt]
	\hline
	\end{tabular}
	  \label{sequences_SEIR-HCD}
	\end{center}
	\end{table}
\end{small}

В таблице~\ref{sequences_SEIR-HCD} приведены последовательности параметров чувствительности, полученные двумя методами. После анализа идентифицируемости можно заключить, что наименее чувствительными (более идентифицируемыми) параметрами модели к вариациям в данных (погрешностям) являются $\alpha_E$, $\kappa$ и $\mu$, другими словами, эти параметры более устойчиво определяются при решении обратной задачи~(\ref{model_COVID})-(\ref{IP_data}). Наиболее чувствительными (менее идентифицируемыми) к ошибкам в измерениях являются параметры $\alpha_I$, $\varepsilon_{CH}$ и $\beta$, то есть необходимо разработать алгоритм регуляризации, позволяющий контролировать качество определения чувствительных параметров.

\section[Математическая модель SEIR-D]{Математическая модель SEIR-D}

В рамках модели SEIR-D распространения коронавируса COVID-19 описывается системой из 5 нелинейных обыкновенных дифференциальных уравнений на отрезке $t\in [t_0,T]$~\cite{Sameni_2020} (схема модели приведена на рис.~\ref{ris:SEIR-HCD_model_scheme} справа):
\begin{eqnarray}\label{model_2}
\left\{\begin{array}{ll}
\dfrac{dS}{dt} = - c(t-\tau) \left(\dfrac{\textcolor{red}{\alpha_I} S(t)I(t)}{N(t)} + \dfrac{\textcolor{red}{\alpha_E} S(t)E(t)}{N(t)}\right) + \gamma R(t),\\[7pt]
\dfrac{dE}{dt} = c(t-\tau) \left(\dfrac{\textcolor{red}{\alpha_I} S(t)I(t)}{N(t)} + \dfrac{\textcolor{red}{\alpha_E} S(t)E(t)}{N(t)}\right) - (\textcolor{red}{\kappa} + \textcolor{red}{\rho}) E(t),\\[7pt]
\dfrac{dI}{dt} = \textcolor{red}{\kappa} E(t) - \textcolor{red}{\beta}I(t)-\textcolor{red}{\mu}I(t),\\[7pt]
\dfrac{dR}{dt} = \textcolor{red}{\beta} I(t) + \textcolor{red}{\rho} E(t) - \gamma R(t),\\[7pt]
\dfrac{dD}{dt} = \textcolor{red}{\mu} I(t);\\[7pt]
\end{array}\right.
\end{eqnarray}

Здесь 
\begin{itemize}
\item $S$~-- восприимчивые (незараженные) индивидуумы;
    \item $E$~-- зараженные индивидуумы без симптомов;
    \item $I$~-- инфицированные индивидуумы с симптомами;
    \item $R$~-- вылеченные индивидуумы;
    \item $D$~-- летальные случаи заболевания среди населения.
    \item $N$ = $S + E + I + R + D$~-- вся популяция.
\end{itemize}

Функция, использующая ограничения на передвижения граждан:
\begin{equation*}
    c(t) = 1 + \textcolor{red}{c^{isol}}\left(1 - \dfrac{2}{5}a(t)\right) , \quad c(t) \in (0,2).
\end{equation*}

Начальные данные:
\begin{equation}\label{model_2_init}
    S(t_0) = S_0, \quad E(t_0) = \textcolor{red}{E_0}, \quad I(t_0) = I_0, \quad R(t_0) = \textcolor{red}{R_0}, \quad D(t_0) = D_0.
\end{equation}

Описание параметров и интервалы поиска приведены в Таблице~\ref{TAB:parameter_info}.

\begin{table}[H]
\caption{Описание параметров моделей~(\ref{model_COVID})-(\ref{init_data}) и (\ref{model_2})-(\ref{model_2_init}) и их пределы.}
\label{TAB:parameter_info}
\begin{tabular}{lp{10cm}p{2.3cm}p{2.1cm}}
\hline\noalign{\smallskip}
Символ & Описание & \multicolumn{2}{c}{Пределы}\\
\multicolumn{2}{c}{} & SEIR-HCD & SEIR-D\\
\noalign{\smallskip}\hline\noalign{\smallskip}
$a(t)$ & индекс самоизоляции по данным Яндекс & \multicolumn{2}{c}{(0, 5)} \\
$\tau$ & латентный период (характеризует запаздывание выделения вирионов или заразности) & \multicolumn{2}{c}{2 дня}\\
$\alpha_I$ & Параметр заражения между инфицированным и восприимчивым населением, который связан с контагиозностью вируса и социальными факторами & \multicolumn{2}{c}{(0, 1)} \\
$\alpha_E$ & Параметр заражения между бессимптомной и восприимчивой группами населения ($\alpha_E >> \alpha_I$) & \multicolumn{2}{c}{(0, 1)} \\
$\kappa$ & Частота появления симптомов в открытых случаях, что приводит к переходу от бессимптомной к инфицированной популяции & \multicolumn{2}{c}{(0, 1)} \\
$\rho$ & Скорость восстановления выявленных случаев (случаи, которые выявлены, но выздоравливают без каких-либо симптомов) & 0  & (0, 1) \\
$\beta$ & Скорость выздоровления зараженных случаев & \multicolumn{2}{c}{(0, 1)} \\
$\gamma$ & Скорость повторного заражения. Этот параметр является обратной величиной уровня иммунитета вируса (0~-- устойчивый иммунитет, 0.001~-- вероятность повторного заражения) & \multicolumn{2}{c}{0}\\
$\nu$ & Доля госпитализированных случаев с тяжелым протеканием заболевания & (0, 1)    &   ~--  \\
$\varepsilon_{HR}$ & Вероятность выздоровления индивидуумов, находящихся в тяжелом состоянии & 0.225    &   ~-- \\
$\varepsilon_{HC}$ & Доля госпитализированных случаев, находящихся в критическом состоянии и требующих подключения аппарата ИВЛ & 0.025   &   ~-- \\
$\varepsilon_{CH}$ & Вероятность отключения аппарата ИВЛ у пациента & (0, 1)   &   ~-- \\
$c^{isol}$ & Коэффициент влияния индекса самоизоляции на заражаемость & ~--  &   (0, 1) \\
$\mu$ & Смертность в результате COVID-19 & ($10^{-4}$, $10^{-1}$)   &   (0, 0.1) \\
$E_0$ & Начальное количество бессимптомных инфицированных & (1, 800) & (0, 600)\\
$R_0$ & Начальное количество вылеченных индивидуумов & ~-- & (0, 600)\\
\noalign{\smallskip}\hline
\end{tabular}
\end{table}

\subsection{Постановка обратной задачи}

Дополнительная информация:
\begin{itemize}
    \item Количество выявленных инфицированных за сутки $f_k$, $k = 1, \dots, K$,
    \item Количество умерших за сутки $g_k$ , $k = 1, \dots, K$.
\end{itemize}

В модели \eqref{model_2} это означает
\begin{equation}\label{model_2_data}
    0.58 f_k = \kappa E_{k-1}, \quad g_k = D(t_k) - D(t_{k-1}), \quad t_k\in(t_0,T),\, k=1,\ldots, K.
\end{equation}

Информации о реальном количестве бессимптомных инфицированных очень мало, но можно предполагать, что за день выявляют почти всех инфицированных с симптомами. \\
В среднем 58 \% среди выявленных за день ($f_k$) --- инфицированные с симптомами ($I$). \\
Выявленные в день $k$ инфицированные с симптомами ($0.58 f_k$), это некоторая доля инфицированных без симптомов в день $k-1$ ($\kappa E(t_{k-1}))$.

Неизвестные параметры модели: \textcolor{red}{$q = (\alpha_E, \alpha_I, \kappa, \rho, \beta, \mu, c^{isol}, E_0, R_0)$} $\in \mathbb{R}^{9}$.

\underline{Обратная задача} \eqref{model_2}-\eqref{model_2_data} состоит в определении вектора параметров $q$ по дополнительной информации \eqref{model_2_data}.

Обратная задача \eqref{model_2}-\eqref{model_2_data} сводится к задаче минимизации функционала: 
\begin{equation}\label{J_2}
    J(q) = \sum_{k=1}^K \Big[ w_1 (\kappa E(t_{k-1};q) - 0.58 f_k)^2 + w_2 (D(t_k;q) - D(t_{k-1};q) - g_k)^2 \Big].
\end{equation}

Веса в функционале \eqref{J_2} выбираются следующим образом:
$$ 
w_1 = \sum_{k=1}^K \frac{f_k}{K}, \qquad  w_2 = \sum_{k=1}^K \frac{g_k}{K}.
$$ 

\section{Численные эксперименты и результаты прогнозирования}
В данном разделе приведено сравнение численных результатов в случае использования обратных задач~(\ref{model_COVID})-(\ref{IP_data}) для SEIR-HCD модели и (\ref{model_2})-(\ref{model_2_data}) для SEIR-D модели для г. Москвы и Новосибирской области. Применяются следующие критерии сравнения:
\begin{enumerate}
\itemsep0em 
    \item Точность предсказания пика эпидемии в г. Москва (дата и количество выявленных инфицированных).
    \item Точность прогноза распространения заболевания в г. Москва по данным о ежедневных выявленных случаях $f_k$ и смертности $g_k$, $k=1,\ldots, K$.
    \item Точность обработки исторических данных ежедневных выявленных случаев $f_k$ и смертности $g_k$ в г. Москва.
    \item Требуемые вычислительные ресурсы (время работы алгоритмов, вычислительные мощности).
    \item Сопоставление необходимой априорной информации для ограничений искомых параметров (см. Таблицу~\ref{TAB:parameter_info}, 3 и 4 колонки).
\end{enumerate}

\subsection{Обработка данных}
Индекс самоизоляции $a(t)$ по Москве и Новосибирской области был учтен из \href{https://yandex.ru/maps/covid19/isolation}{карт Яндекс}. Коэффициент тестирования вычислялся с использованием \href{https://стопкоронавирус.рф/}{статистической информации} из открытых источников по формуле: 
\[c^{test}_k = \dfrac{T_k}{Z_k},\]
где $T_k$~-- количество сделанных в регионе тестов в день $k$, полученное умножением количества тестов, сделанных в стране в день $k$, на долю населения региона в населении страны, $Z_k$~-- количество незараженных индивидуумов в регионе ко дню $k$, посчитанное как разность числа жителей региона и общего числа заболевших и умерших за весь период наблюдения до дня $k$ включительно (данные о вылеченных не использовались). Для прогнозирования $T_k$ и $Z_k$ линейно экстраполировались на дальнейший период времени.

\subsection{Алгоритмы решения обратных задач}

\subsubsection{Для модели SEIR-HCD}
Для численного решения обратной задачи~(\ref{model_COVID})-(\ref{IP_data}) использовалась следующая последовательность шагов: 
\begin{enumerate}
\itemsep0em 
    \item Обработка данных $f_k$, $g_k$, $c^{test}_k$, $a(t_k)$, $k=1,...,K$ для временных периодов a)-d). 
    \item Определение границ параметров для искомого вектора $q$ (см. Таблицу~\ref{TAB:parameter_info}, колонку 3).
    \item Применение метода дифференциальной эволюции посредством модуля \textit{scipy.optimize.differential\_evolution} библиотеки \textit{SciPy} для определения минимума функционала~(\ref{func_SEIR-HCD}) и получения оптимального вектора параметров $q^*$ математической модели (\ref{model_COVID}), (\ref{init_data}):
    \begin{enumerate}
    \itemsep0em 
        \item[3.1.] Решение прямой задачи~(\ref{model_COVID}), (\ref{init_data}) и вычисление значения функционала~(\ref{func_SEIR-HCD}) на каждой итерации метода. Для решения прямой задачи~(\ref{model_COVID}), (\ref{init_data}) использовался модуль \textit{scipy.integrate.odeint} библиотеки \textit{SciPy}. 
        \item[3.2.] Определение оптимального вектора параметров $q^*$.
    \end{enumerate}
    \item Решение прямой задачи~(\ref{model_COVID}), (\ref{init_data}) для временных периодов a)--d) с использованием найденных оптимальных параметров из $q^*$ и построение прогнозов для соответствующих временных периодов a)--d). Количество выявленных за сутки инфицированных $\hat{f}_k$ вычисляется по формуле $\hat{f}_k = c^{test}_k(I(t_{k}) + E(t_{k}))$.
\end{enumerate}

\subsubsection{Для модели SEIR-D}
Вычисления проводились с помощью языка программирования Python. Для численного решения обратной задачи~(\ref{model_2})---(\ref{model_2_data}) использовалась следующая последовательность шагов: 
\begin{enumerate}
\itemsep0em 
    \item Обработка данных $f_k$, $g_k$, $a(t_k)$, $k=1,...,K$, для временных периодов a)-d).
    \item Определение границ параметров для искомого вектора $q$ (см. Таблицу~\ref{TAB:parameter_info}, колонку 4).
    \item Определение минимума функционала~(\ref{J_2}) и получение оптимальных векторов параметров $q^*$ математической модели (\ref{model_2}), (\ref{model_2_init}) с помощью четырёх различных методов:
        \begin{enumerate}
    \itemsep0em 
        \item[1)] метод дифференциальной эволюции (модуль \textit{scipy.optimize.differential\_evolution}), 
        \item[2)] метод имитации отжига (модуль \textit{scipy.optimize.dual\_annealing}),
        \item[3)] генетический алгоритм (\href{https://pypi.org/project/geneticalgorithm/}{библиотека}),
        \item[4)] метод роя частиц (\href{https://pythonhosted.org/pyswarm/}{библиотека}).
        \end{enumerate}
    \item Из полученных методами 1)--4) векторов $q^*$ выбирается лучший в смысле функционала \eqref{J_2}:
        \begin{enumerate}
    \itemsep0em 
        \item[4.1.] Решение прямой задачи~(\ref{model_2}), (\ref{model_2_init}) и вычисление значения функционала~(\ref{J_2}) на каждой итерации метода. Для решения прямой задачи~(\ref{model_2}), (\ref{model_2_init}) использовался модуль \textit{scipy.integrate.ode\_ivp} библиотеки \textit{SciPy}, метод \textit{RK45}. 
        \item[4.2.] Определение оптимального вектора параметров $q^*$ математической модели~(\ref{model_2}), (\ref{model_2_init}).
        \end{enumerate}
    \item Решение прямой задачи~(\ref{model_2}), (\ref{model_2_init}) для временных периодов a)--d) с использованием найденных оптимальных параметров из $q^*$ и построение прогнозов для соответствующих временных периодов a)--d). Количество выявленных за сутки инфицированных $\hat{f}_k$ вычисляется по формуле $\hat{f}_k = \dfrac{\kappa E(t_{k-1})}{0.58}$.
\end{enumerate}

\subsection{Численные расчеты для г. Москва}
В качестве начальных данных распространения коронавируса COVID-19 в городе Москва ($N_0 = 11~514~330$ человек) в рамках математической модели SEIR-HCD~(\ref{model_COVID}) была использована \href{https://xn--80aesfpebagmfblc0a.xn--p1ai/#}{статистическая информация} за 23 марта 2020 года:
\[S_0 = 11~514~241-q_8,\, E_0 = q_8,\, I_0 = 71,\, R_0 = 9,\, H_0 = 8,\, C_0 = 1,\, D_0 = 0.\]

В модели SEIR-D~(\ref{model_2}) использовались следующие начальные данные:  
\[S_0 = 11~514~330-I_0-q_8-q_9,\, E_0 = q_8,\, I_0 = 71,\, R_0 = q_9,\, D_0 = 0.\]

Данные обратных задач были получены из открытых источников для г. Москвы для временных периодов:
\begin{enumerate}
    \itemsep0em 
        \item[а)] 23.03.2020 -- 21.05.2020 ($K=60$ дней), \qquad б) 23.03.2020 -- 11.05.2020 ($K=50$ дней),
        \item[в)] 23.03.2020 -- 01.05.2020 ($K=40$ дней), \qquad г) 23.03.2020 -- 21.04.2020 ($K=30$ дней),
        \item[д)] 23.03.2020 -- 15.06.2020 ($K=85$ дней).
\end{enumerate}

Вычисления проводились с помощью языка программирования Python. Прогноз строился для следующих временных периодов в соответствии с данными а)--д), указанными выше:
    \begin{enumerate}
    \itemsep0em 
        \item[а)] 22.05.2020 -- 21.06.2020 (30 дней), \qquad б) 12.05.2020 -- 11.06.2020 (30 дней),
        \item[в)] 02.05.2020 -- 01.06.2020 (30 дней), \qquad г) 22.04.2020 -- 21.05.2020 (30 дней),
        \item[д)] 16.06.2020 -- 20.06.2020 (5 дней).
    \end{enumerate}

\subsubsection{Сравнение результатов}
Было проведено сравнение результатов, полученных по двум математическим моделям типа SEIR-HCD и SEIR-D, параметры которых были уточнены для г. Москвы методами оптимизации (метод дифференциальной эволюции, метод имитации отжига, генетический алгоритм, метод роя частиц). 

\begin{figure}[H]
  \begin{minipage}[h]{0.49\linewidth}
  \center{\includegraphics[width=1\linewidth]{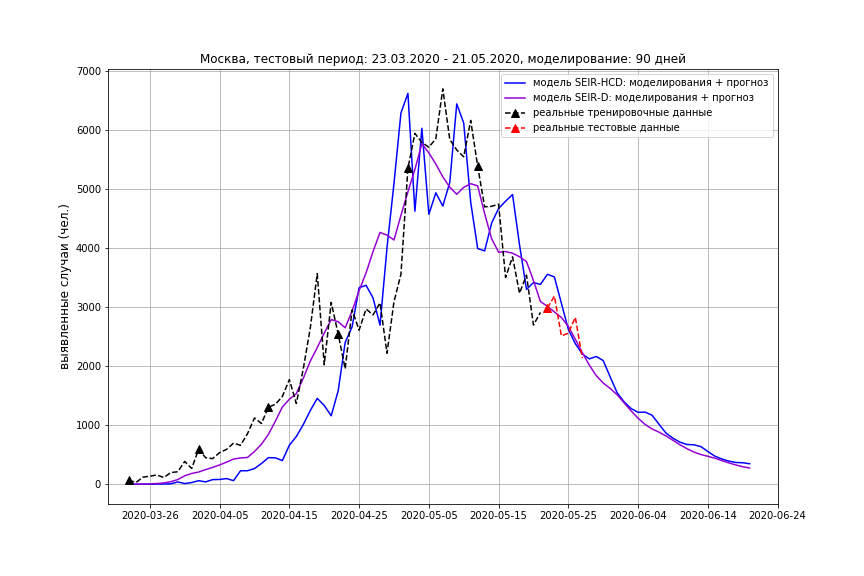}} а) Моделирование по 21.05.2020 (90 дней).\\
  Параметры приведены в табл.~\ref{TAB:parameters_0}.
  \end{minipage}
  \hfill
  \begin{minipage}[h]{0.49\linewidth}
  \center{\includegraphics[width=1\linewidth]{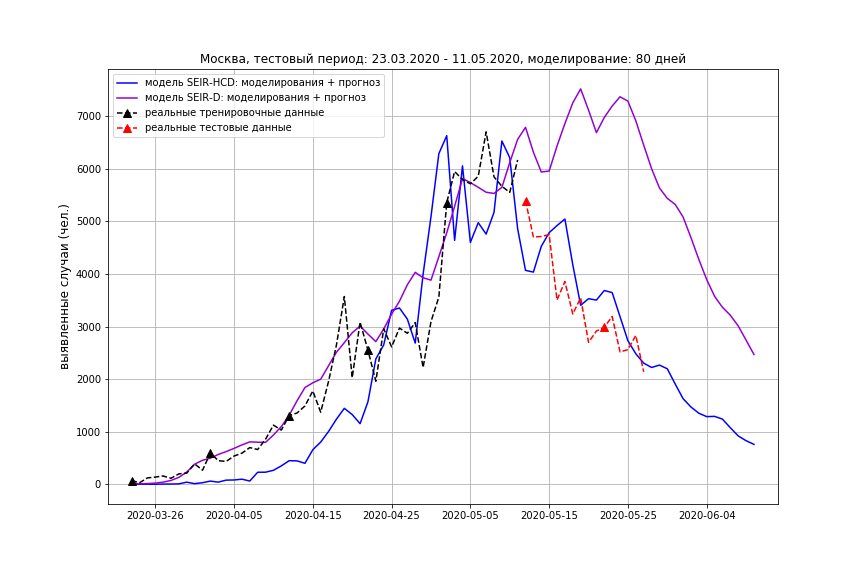}} б) Моделирование по 11.05.2020 (80 дней).\\
  Параметры приведены в табл.~\ref{TAB:parameters_10}.
  \end{minipage}
  \vfill
  \begin{minipage}[h]{0.49\linewidth}
  \center{\includegraphics[width=1\linewidth]{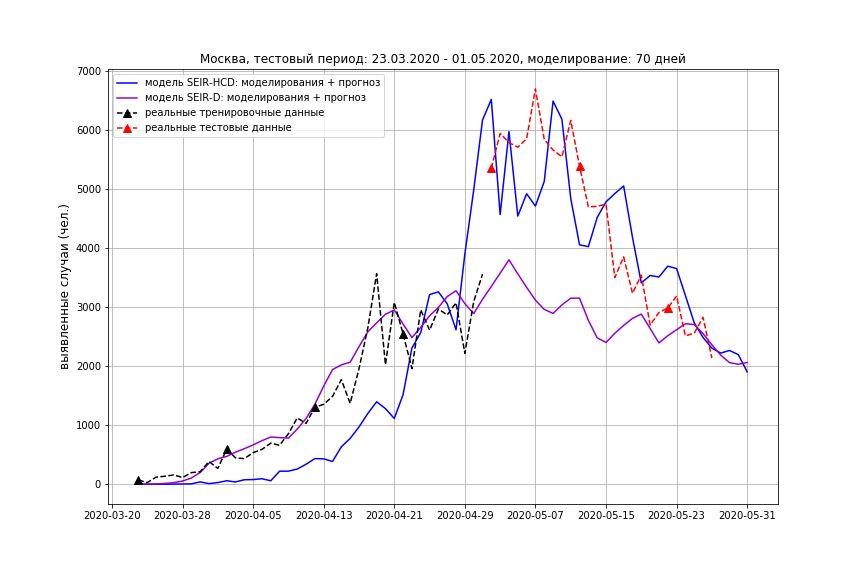}} в) Моделирование по 01.05.2020 (70 дней).\\
  Параметры приведены в табл.~\ref{TAB:parameters_20}.
  \end{minipage}
  \hfill
  \begin{minipage}[h]{0.49\linewidth}
  \center{\includegraphics[width=1\linewidth]{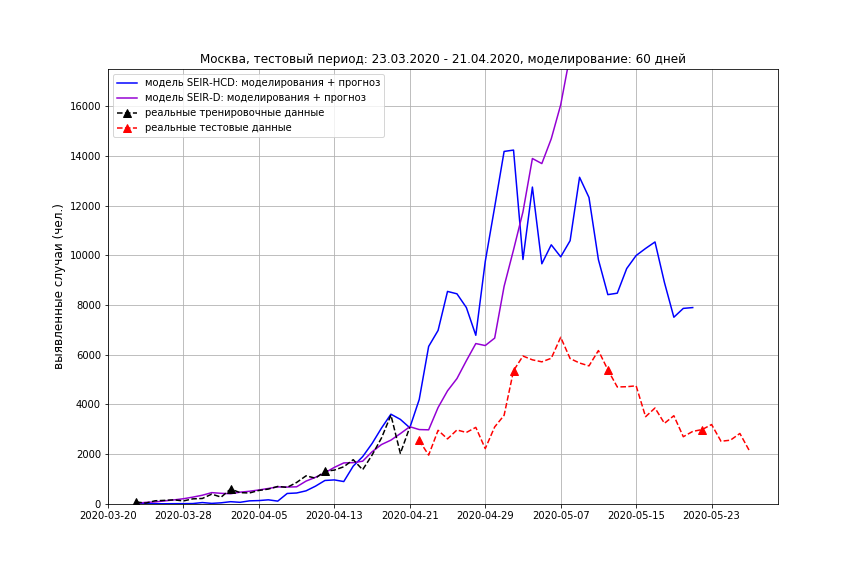}} г) Моделирование по 21.04.2020 (60 дней).\\
  Параметры приведены в табл.~\ref{TAB:parameters_30}.
  \end{minipage}
  \vfill
  \begin{minipage}[h]{0.49\linewidth}
  \center{\includegraphics[width=1\linewidth]{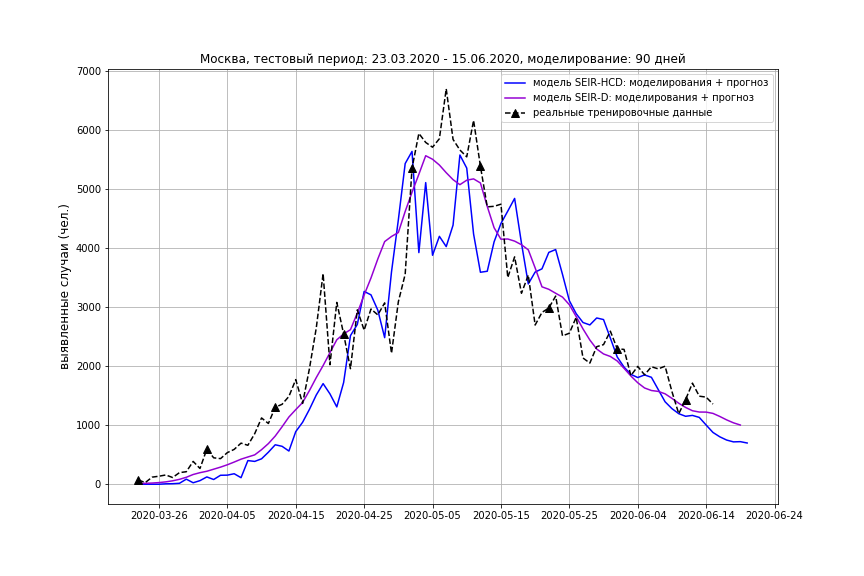}} д) Моделирование по 15.06.2020 (90 дней).\\
  Параметры приведены в табл.~\ref{TAB:parameters_00}.
  \end{minipage}
  \caption{Количество выявленных случаев в г. Москва в случае измерений с 23.03.2020 до указанных на графиках периодов моделирования 90, 80, 70, 60 и 90 дней, соответственно.\\ Пунктирная черная линия -- реальные данные $f_k$, использованные при решении обратных задач, пунктирная красная линия -- реальные данные до 27.05.2020, не использованные для решения обратных задач. Синяя линия -- решение для математической модели SEIR-HCD, фиолетовая линия -- решение для математической модели SEIR-D. Значения восстановленных параметров для математических моделей приведены в таблицах~\ref{TAB:parameters_0}-\ref{TAB:parameters_00}.}
  \label{ris:Moscow_Tested_compare}
\end{figure}

На рисунке~\ref{ris:Moscow_Tested_compare} представлено количество выявленных инфицированных для двух моделей, верифицированных на реальных данных до 27 мая 2020 года для временных периодов а)--d). Отметим, что даже при использовании меньшего количества статистических данных прогноз распространения эпидемии более точен в случае математической модели SEIR-HCD (синяя линия), т.е. с хорошей точностью построен прогноз для количества выявленных случаев в Москве (см. отношение результата моделирования SEIR-HCD, обозначенного синей линией, к неиспользованным в решении обратной задачи статистическим данным, обозначенным пунктирной красной линией). Математическая модель SEIR-D имеет бóльшую погрешность в прогнозировании количества выявленных случаев, однако точнее описывает исторические данные (см. отношение результата моделирования SEIR-D, обозначенного фиолетовой линией, к использованным в решении обратной задачи статистическим данным, обозначенным пунктирной черной линией). В случае наиболее полных данных измерений с 23 марта по 15 июня 2020 года (85 дней) обе математические модели дают схожие результаты (Рис.~\ref{ris:Moscow_Tested_compare}д). Более детальное сравнение для временных периодов а)--г) в рамках критериев 1-5 приведено в Таблицах~\ref{TAB:compare_21-05-2020}-\ref{TAB:compare_21-04-2020} (с учетом качества прогнозирования).

Восстановленные параметры для моделей SEIR-HCD и SEIR-D приведены в Таблицах~\ref{TAB:parameters_0}-\ref{TAB:parameters_00}. Анализируя динамику поведения параметров для математической модели SEIR-HCD при разном наборе данных, отметим, что параметры $\varepsilon_{CH}$ и $\beta$ меняются в больших диапазонах в отличие от остальных параметров, о чем было отмечено в рамках проведенного анализа чувствительности (раздел~\ref{sec_identifiability}). Таким образом, при решении обратной задачи необходимо применять регуляризацию, например, в виде статистических ограничений, что было использовано в расчетах.

\begin{tiny}
\begin{table}[!ht]
\caption{Восстановленные параметры для периода моделирования 23.03.20-21.05.20.}
\label{TAB:parameters_0}
\begin{tabular}{llllllllllll}
\hline\noalign{\smallskip}
Модель & $\alpha_E$ & $\alpha_I$ & $\kappa$ & $\rho$ & $\beta$ & $\nu$ & $\varepsilon_{CH}$ & $\mu$ & $c^{isol}$ & $E_0$ & $R_0$\\
\noalign{\smallskip}\hline\noalign{\smallskip}
SEIR-HCD & 0.995 & 0.499 & 0.28 & -- & 0.146 & 0.06 & 0.003 & 0.0076 & -- & 795 & --\\
SEIR-D & 0.510 & 0.618 & 0.002 & 0.187 & 0.230 & -- & -- & 0.0052 & 0.9 & 783 & 187\\
\noalign{\smallskip}\hline
\end{tabular}
\end{table}
\end{tiny}

\begin{tiny}
\begin{table}[!ht]
\caption{Восстановленные параметры для периода моделирования 23.03.20-11.05.20.}
\label{TAB:parameters_10}
\begin{tabular}{llllllllllll}
\hline\noalign{\smallskip}
Модель & $\alpha_E$ & $\alpha_I$ & $\kappa$ & $\rho$ & $\beta$ & $\nu$ & $\varepsilon_{CH}$ & $\mu$ & $c^{isol}$ & $E_0$ & $R_0$\\
\noalign{\smallskip}\hline\noalign{\smallskip}
SEIR-HCD & 0.999 & 0.5 & 0.29 & -- & 0.16 & 0.037 & 0.001 & 0.011 & -- & 798 & --\\
SEIR-D & 0.874 & 0.669 & 0.037 & 0.66 & 0.97 & -- & -- & 0.0163 & 0.346 & 598 & 48\\
\noalign{\smallskip}\hline
\end{tabular}
\end{table}
\end{tiny}

\begin{tiny}
\begin{table}[!ht]
\caption{Восстановленные параметры для периода моделирования 23.03.20-01.05.20.}
\label{TAB:parameters_20}
\begin{tabular}{llllllllllll}
\hline\noalign{\smallskip}
Модель & $\alpha_E$ & $\alpha_I$ & $\kappa$ & $\rho$ & $\beta$ & $\nu$ & $\varepsilon_{CH}$ & $\mu$ & $c^{isol}$ & $E_0$ & $R_0$\\
\noalign{\smallskip}\hline\noalign{\smallskip}
SEIR-HCD & 0.994 & 0.491 & 0.28 & -- & 0.165 & 0.04 & 0.0065 & 0.011 & -- & 794 & --\\
SEIR-D & 0.849 & 0.716 & 0.011 & 0.58 & 0.36 & -- & -- & 0.0089 & 0.56 & 586 & 146\\
\noalign{\smallskip}\hline
\end{tabular}
\end{table}
\end{tiny}

\begin{tiny}
\begin{table}[!ht]
\caption{Восстановленные параметры для периода моделирования 23.03.20-21.04.20.}
\label{TAB:parameters_30}
\begin{tabular}{llllllllllll}
\hline\noalign{\smallskip}
Модель & $\alpha_E$ & $\alpha_I$ & $\kappa$ & $\rho$ & $\beta$ & $\nu$ & $\varepsilon_{CH}$ & $\mu$ & $c^{isol}$ & $E_0$ & $R_0$\\
\noalign{\smallskip}\hline\noalign{\smallskip}
SEIR-HCD & 0.998 & 0.493 & 0.268 & -- & 0.063 & 0.007 & 0.171 & 0.0473 & -- & 798 & --\\
SEIR-D & 0.078 & 0.787 & 0.643 & 0.703 & 0.173 & -- & -- & 0.0039 & 0.63 & 17 & 332\\
\noalign{\smallskip}\hline
\end{tabular}
\end{table}
\end{tiny}

\begin{tiny}
\begin{table}[!ht]
\caption{Восстановленные параметры для периода моделирования 23.03.20-15.06.20.}
\label{TAB:parameters_00}
\begin{tabular}{llllllllllll}
\hline\noalign{\smallskip}
Модель & $\alpha_E$ & $\alpha_I$ & $\kappa$ & $\rho$ & $\beta$ & $\nu$ & $\varepsilon_{CH}$ & $\mu$ & $c^{isol}$ & $E_0$ & $R_0$\\
\noalign{\smallskip}\hline\noalign{\smallskip}
SEIR-HCD & 0.983 & 0.249 & 0.24 & -- & 0.099 & 0.099 & $1.4\cdot 10^{-8}$ & 0.0034 & -- & 2208 & --\\
SEIR-D & 0.999 & 0.046 & 0.079 & 0.706 & 0.029 & -- & -- & 0.0013 & 0.27 & 98 & 6\\
\noalign{\smallskip}\hline
\end{tabular}
\end{table}
\end{tiny}

\begin{figure}[H]
  \begin{minipage}[h]{0.49\linewidth}
  \center{\includegraphics[width=1\linewidth]{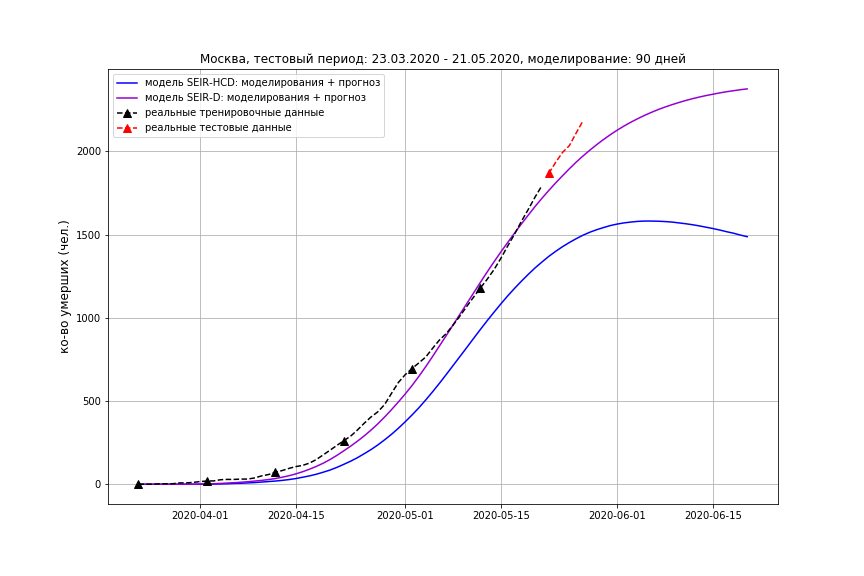}}\\ а) Моделирование по 21.05.2020 (90 дней).\\
  Параметры приведены в табл.~\ref{TAB:parameters_0}.
  \end{minipage}
  \hfill
  \begin{minipage}[h]{0.49\linewidth}
  \center{\includegraphics[width=1\linewidth]{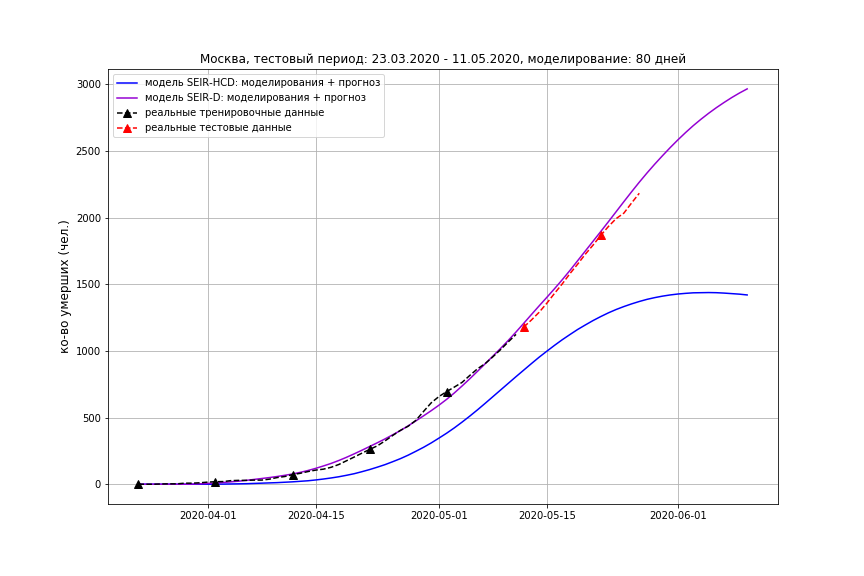}}\\ б) Моделирование по 11.05.2020 (80 дней).\\
  Параметры приведены в табл.~\ref{TAB:parameters_10}.
  \end{minipage}
  \vfill
  \begin{minipage}[h]{0.49\linewidth}
  \center{\includegraphics[width=1\linewidth]{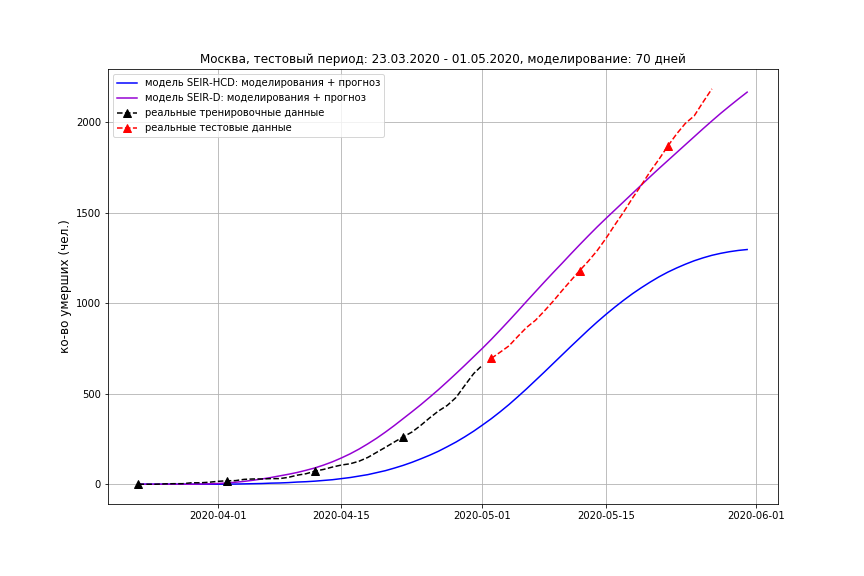}}\\ в) Моделирование по 01.05.2020 (70 дней).\\
  Параметры приведены в табл.~\ref{TAB:parameters_20}.
  \end{minipage}
  \hfill
  \begin{minipage}[h]{0.49\linewidth}
  \center{\includegraphics[width=1\linewidth]{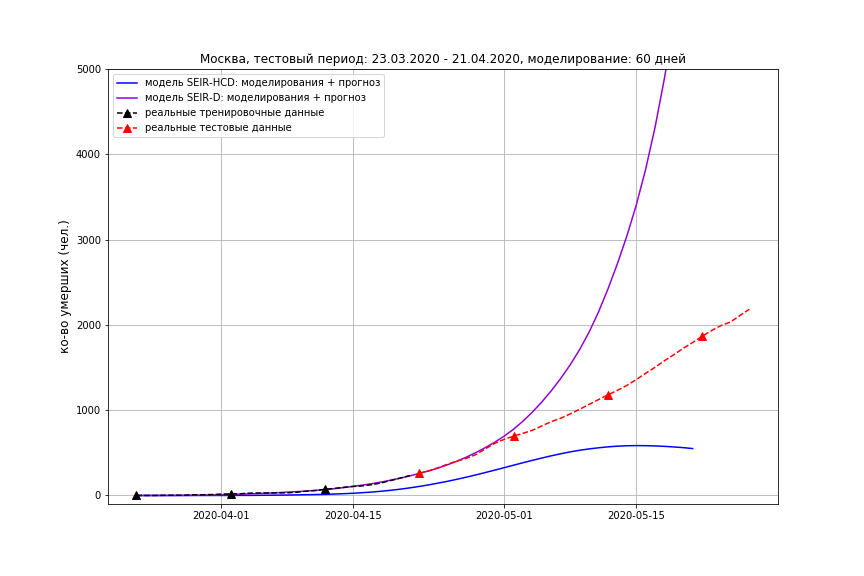}}\\ г) Моделирование по 21.04.2020 (60 дней).\\
  Параметры приведены в табл.~\ref{TAB:parameters_30}.
  \end{minipage}
  \vfill
  \begin{minipage}[h]{0.49\linewidth}
  \center{\includegraphics[width=1\linewidth]{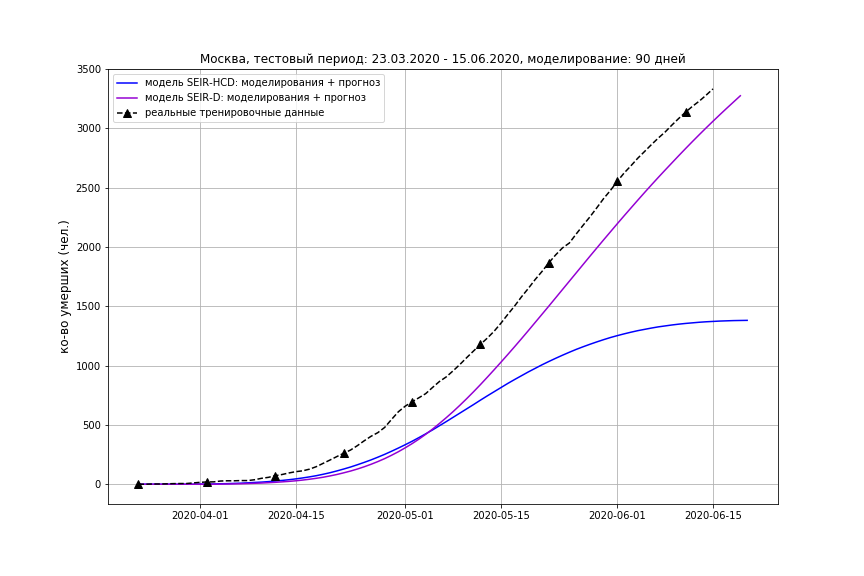}} д) Моделирование по 15.06.2020 (85 дней).\\
  Параметры приведены в табл.~\ref{TAB:parameters_00}.
  \end{minipage}
  \caption{Кумулятивная смертность в г. Москва в случае измерений с 23.03.2020 до указанных на графиках периодов моделирования 90, 80, 70, 60 и 85 дней, соответственно.\\ Пунктирная черная линия -- реальные данные $g_k$, использованные при решении обратных задач, пунктирная красная линия -- реальные данные до 27.05.2020, не использованные для решения обратных задач. Синяя линия -- решение для математической модели SEIR-HCD, фиолетовая линия -- решение для математической модели SEIR-D. Значения восстановленных параметров для математических моделей приведены в таблицах~\ref{TAB:parameters_0}-\ref{TAB:parameters_00}.} 
  \label{ris:Moscow_Death_compare}
\end{figure}

Анализируя результаты расчетов для группы $D(t)$, представленные на Рис.~\ref{ris:Moscow_Death_compare}, можно заключить, что прогнозирование посредством уточнения коэффициентов математической модели SEIR-D (фиолетовая линия) значительно лучше, чем в случае прогнозирования посредством уточнения коэффициентов математической модели SEIR-HCD (синяя линия), то же касается и качественного описания исторических данных $g_k$. Исключение составляет период с меньшей статистикой (до 21 апреля), представленный на рисунке~\ref{ris:Moscow_Death_compare}г. Убывание кривой математической модели SEIR-HCD (синяя линяя) на графике~\ref{ris:Moscow_Death_compare}а обусловлено ошибками округления (в данном случае имеется в виду отсутствие смертности в период постоянства и убывания кривой).

В Таблицах~\ref{TAB:compare_21-05-2020}-\ref{TAB:compare_21-04-2020} приведены варианты сравнения результатов моделирования $\hat{y}$ с реальными данными $y$ для данных типа а)--г) согласно следующим мерам с учетом длины временного периода $M$:
\begin{itemize}
\item среднеквадратичная ошибка: \href{https://scikit-learn.org/stable/modules/generated/sklearn.metrics.mean_squared_error.html#sklearn.metrics.mean_squared_error}{MSE}$(y,\hat{y}) =  \sum\limits_{k=1}^{M} \dfrac{(y_k - \hat{y}_k)^2}{M}$;
    \item среднее отклонение: \href{https://scikit-learn.org/stable/modules/generated/sklearn.metrics.mean_absolute_error.html#sklearn.metrics.mean_absolute_error}{MAE}$(y,\hat{y}) =  \sum\limits_{k=1}^{M} \dfrac{|y_k - \hat{y}_k|}{M}$.
\end{itemize}

Введем обозначения:\\
$Err^{MAE}_{fore}=$MAE$(f,\hat{f})$ и $Err^{MSE}_{fore}=$MSE$(f,\hat{f})$ -- ошибки при соответствующем тестовом периоде $N$, данные из которого не использовались при решении обратных задач (на графиках~\ref{ris:Moscow_Tested_compare} и~\ref{ris:Moscow_Death_compare} данные обозначены пунктирной красной линией),\\
$Err^{MAE}_{desc}=$MAE$(f,\hat{f})$ и $Err^{MSE}_{desc}=$MSE$(f,\hat{f})$ -- ошибки при соответствующем периоде обучения $K$, данные $f_k$ и $g_k$ из которого использовались для решения обратных задач (на графиках~\ref{ris:Moscow_Tested_compare} и~\ref{ris:Moscow_Death_compare} данные обозначены пунктирной черной линией).

Для проверки критерия 1 введем $\delta_t$~-- отклонение даты пика эпидемии в модели от даты реального пика - 7 мая 2020 года (в днях), $\delta_f$~-- отклонение величины пика эпидемии в модели от реальной величины пика - в 6703 выявленных случаев (в количестве людей).

\begin{table}[!ht]
\caption{Сравнение моделей для данных 23.03.20-21.05.20, $K=60$, $N=6$.}
\label{TAB:compare_21-05-2020}
\begin{tabular}{lp{1cm}p{1cm}p{2cm}p{2cm}p{2cm}p{2cm}}
\hline\noalign{\smallskip}
Модель & \multicolumn{2}{c}{Критерий 1} & \multicolumn{2}{c}{Критерий 2} & \multicolumn{2}{c}{Критерий 3}\\
{} & $\delta_t$ & $\delta_f$ & $Err^{MAE}_{fore}$ & $Err^{MSE}_{fore}$ & $Err^{MAE}_{desc}$ & $Err^{MSE}_{desc}$\\
\noalign{\smallskip}\hline\noalign{\smallskip}
SEIR-HCD & 2 & 254 & 338 & 157690 & 769 & 931618\\
SEIR-D & 3 & 930 & 198 & 54977 & 454 & 365087\\
\noalign{\smallskip}\hline
\end{tabular}
\end{table}

\begin{table}[!ht]
\caption{Сравнение моделей для данных 23.03.20-11.05.20, $K=50$, $N=16$.}
\label{TAB:compare_11-05-2020}
\begin{tabular}{lp{1cm}p{1cm}p{2cm}p{2cm}p{2cm}p{2cm}}
\hline\noalign{\smallskip}
Модель & \multicolumn{2}{c}{Критерий 1} & \multicolumn{2}{c}{Критерий 2} & \multicolumn{2}{c}{Критерий 3}\\
{} & $\delta_t$ & $\delta_f$ & $Err^{MAE}_{fore}$ & $Err^{MSE}_{fore}$ & $Err^{MAE}_{desc}$ & $Err^{MSE}_{desc}$\\
\noalign{\smallskip}\hline\noalign{\smallskip}
SEIR-HCD & 2 & 174 & 614 & 556795 & 780 & 974866\\
SEIR-D & 12 & 817 & 3346 & 12745883 & 338 & 247688\\
\noalign{\smallskip}\hline
\end{tabular}
\end{table}

\begin{table}[!ht]
\caption{Сравнение моделей для данных 23.03.20-01.05.20, $K=40$, $N=26$.}
\label{TAB:compare_01-05-2020}
\begin{tabular}{lp{1cm}p{1cm}p{2cm}p{2cm}p{2cm}p{2cm}}
\hline\noalign{\smallskip}
Модель & \multicolumn{2}{c}{Критерий 1} & \multicolumn{2}{c}{Критерий 2} & \multicolumn{2}{c}{Критерий 3}\\
{} & $\delta_t$ & $\delta_f$ & $Err^{MAE}_{fore}$ & $Err^{MSE}_{fore}$ & $Err^{MAE}_{desc}$ & $Err^{MSE}_{desc}$\\
\noalign{\smallskip}\hline\noalign{\smallskip}
SEIR-HCD & 2 & 207 & 779 & 842300 & 714 & 889008\\
SEIR-D & 3 & 2898 & 1546 & 3531484 & 233 & 105670\\
\noalign{\smallskip}\hline
\end{tabular}
\end{table}

\begin{table}[!ht]
\caption{Сравнение моделей для данных 23.03.20-21.04.20, $K=30$, $N=36$.}
\label{TAB:compare_21-04-2020}
\begin{tabular}{lp{1cm}p{1cm}p{2cm}p{2cm}p{2cm}p{2cm}}
\hline\noalign{\smallskip}
Модель & \multicolumn{2}{c}{Критерий 1} & \multicolumn{2}{c}{Критерий 2} & \multicolumn{2}{c}{Критерий 3}\\
{} & $\delta_t$ & $\delta_f$ & $Err^{MAE}_{fore}$ & $Err^{MSE}_{fore}$ & $Err^{MAE}_{desc}$ & $Err^{MSE}_{desc}$\\
\noalign{\smallskip}\hline\noalign{\smallskip}
SEIR-HCD & 5 & 7537 & 5242 & 30788912 & 342 & 183419\\
SEIR-D & 31 & 348120 & 48996 & 6350947086 & 148 & 69867\\
\noalign{\smallskip}\hline
\end{tabular}
\end{table}

\subsection{Численные расчеты для Новосибирской области}
В качестве начальных данных распространения коронавируса COVID-19 в Новосибирской области ($N_0 = 2~798~170$ человек) для математической модели SEIR-HCD была взята статистическая информация за 23 марта 2020 года:
\[S_0 = 2~798~170 - q_8,\, E_0 = q_8,\, I_0 = 0,\, R_0 = 0,\, H_0 = 0,\, C_0 = 0,\, D_0 = 0,\]
а для математической модели SEIR-D в следующем виде:
\[S_0 = 2~798~170-q_8-q_9,\, E_0 = q_8,\, I_0 = 0,\, R_0 = q_9,\, D_0 = 0.\]

\begin{figure}[H]
  \begin{minipage}[h]{0.49\linewidth}
  \center{\includegraphics[width=1\linewidth]{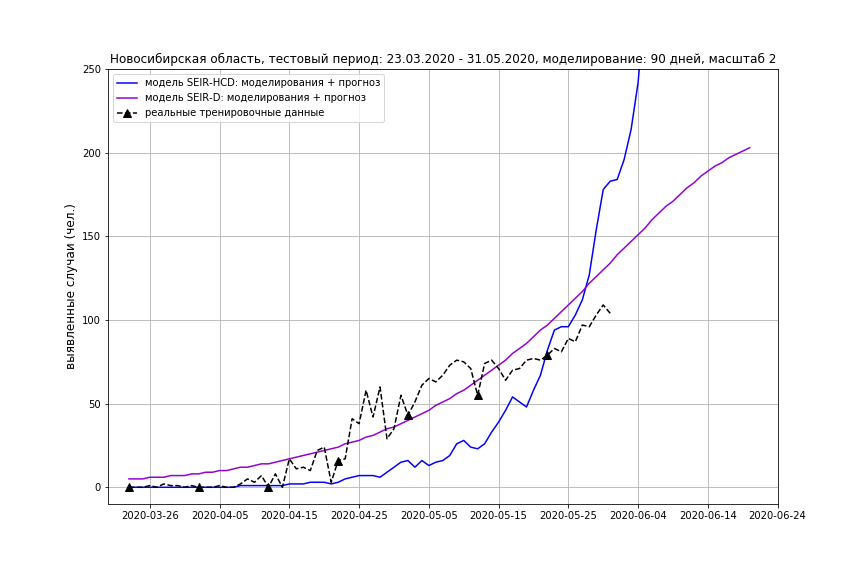}} а) Моделирование по 31.05.2020.\\
  Параметры приведены в табл.~\ref{TAB:Nsk_parameters_0}.
  \end{minipage}
  \hfill
  \begin{minipage}[h]{0.49\linewidth}
  \center{\includegraphics[width=1\linewidth]{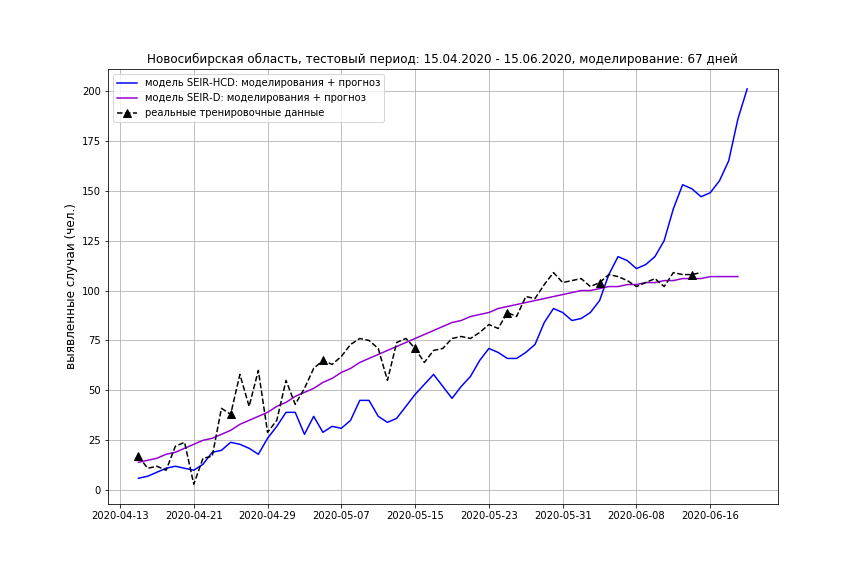}} б) Моделирование по 15.06.2020.\\
  Параметры приведены в табл.~\ref{TAB:Nsk_parameters_00}.
  \end{minipage}
  \caption{Количество выявленных случаев COVID-19 в Новосибирской области с 23.03.2020 по 21.06.2020 (90 дней) в случае измерений с 23.03.2020 по а) 31.05.2020 и б) 15.06.2020. Синяя линия -- решение для математической модели SEIR-HCD, фиолетовая линия -- решение для математической модели SEIR-D, пунктирная черная линия -- реальные данные $f_k$, использованные при решении обратных задач. Значения восстановленных параметров для математических моделей приведены в таблицах~\ref{TAB:Nsk_parameters_0}-\ref{TAB:Nsk_parameters_00}.} 
  \label{ris:Nsk_compare_0}
\end{figure}

На рисунке~\ref{ris:Nsk_compare_0} приведены результаты моделирования и прогнозирования количества выявленных случаев в Новосибирской области в случае измерений с 23 марта по 31 мая 2020 года (Рис.~\ref{ris:Nsk_compare_0}a) и по 15 июня 2020 года (Рис.~\ref{ris:Nsk_compare_0}б), полученные с использованием описанных алгоритмов для двух математических моделей, указанных выше. Отметим, что описание реальных данных по выявленным случаям, полученное с помощью математической модели SEIR-D (фиолетовая линия), точнее, чем моделирование с помощью модели SEIR-HCD (синяя линия).

Анализ моделирования смертности в Новосибирской области затруднителен для обеих моделей, так как имеющихся данных недостаточно для качественного прогнозирования смертности в регионе. Отметим, что влияние коэффициента индекса самоизоляции на заражаемость $c^{isol}$ в Новосибирской области для математической модели SEIR-D нулевое, что, фактически, означает отсутствие влияние изоляции на заражаемость в регионе (см. Таблицы~\ref{TAB:Nsk_parameters_0}-\ref{TAB:Nsk_parameters_00}).

\begin{tiny}
\begin{table}[!ht]
\caption{Восстановленные параметры для периода моделирования 23.03.20-31.05.20, Новосибирская область.}
\label{TAB:Nsk_parameters_0}
\begin{tabular}{llllllllllll}
\hline\noalign{\smallskip}
Модель & $\alpha_E$ & $\alpha_I$ & $\kappa$ & $\rho$ & $\beta$ & $\nu$ & $\varepsilon_{CH}$ & $\mu$ & $c^{isol}$ & $E_0$ & $R_0$\\
\noalign{\smallskip}\hline\noalign{\smallskip}
SEIR-HCD & 0.001 & 0.224 & 0.108 & -- & 0.013 & 0.006 & 0.055 & 0.072 & -- & 1001 & --\\
SEIR-D & 0.999 & 0.999 & 0.042 & 0.952 & 0.999 & -- & -- & 0.0188 & 0 & 99 & 24\\
\noalign{\smallskip}\hline
\end{tabular}
\end{table}
\end{tiny}

\begin{tiny}
\begin{table}[!ht]
\caption{Восстановленные параметры для периода моделирования 23.03.20-15.06.20, Новосибирская область.}
\label{TAB:Nsk_parameters_00}
\begin{tabular}{llllllllllll}
\hline\noalign{\smallskip}
Модель & $\alpha_E$ & $\alpha_I$ & $\kappa$ & $\rho$ & $\beta$ & $\nu$ & $\varepsilon_{CH}$ & $\mu$ & $c^{isol}$ & $E_0$ & $R_0$\\
\noalign{\smallskip}\hline\noalign{\smallskip}
SEIR-HCD & 0.219 & 0.0001 & 0.118 & -- & 0.046 & 0.001 & 0.001 & 0.1 & -- & 6412 & --\\
SEIR-D & 0.271 & 0.999 & $1.9\cdot 10^{-5}$ & $10^{-9}$ & 0.007 & -- & -- & 0.0009 & 0 & 93 & 96\\
\noalign{\smallskip}\hline
\end{tabular}
\end{table}
\end{tiny}

\section{Заключение}
Были проанализированы две математические модели SEIR-HCD и SEIR-D, основанные на SEIR-структуре, сформулированы для каждой из них обратные задачи уточнения параметров для г. Москва и Новосиб ирской области, проведен анализ идентифицируемости и чувствительности параметров к ошибкам в статистических данных, а также разработаны алгоритмы решения обратных задач. Построены сценарии развития распространения эпидемии COVID-19 в г. Москва и Новосибирской области для различных временных промежутков измерений и прогнозирования.

Прогнозирование пика развития эпидемии в г. Москве с ошибкой в 2 дня и на 174 выявленных случая меньше фактического было получено с помощью математической модели SEIR-HCD при использовании данных по выявленным случаям и смертности в Москве с 23 марта 2020 года по 11 мая 2020 года. Также прогнозирование сценария развития эпидемии (количество новых выявленных случаев за день) с наименьшей ошибкой было получено с помощью математической модели SEIR-HCD во всех случаях использования данных (30, 40, 50 дней статистики), кроме наибольшего периода (60 дней статистики), при котором меньшую ошибку продемонстрировала математическая модель SEIR-D. По-видимому, использование более грубой математической модели (с меньшим количеством однородных групп) оправдано в случае достаточно большого количества статистических данных и небольшого периода прогнозирования. Действительно, математическая модель SEIR-D наилучшим образом приближает исторические данные по выявленным случаям и смертности для г. Москвы.

Моделирование и прогнозирование развития коронавирусной инфекции в Новосибирской области продемонстрировано для двух наборов измерений: с 23.03.2020 по 31.05.2020 и по 15.06.2020 ввиду небольшой статистики вплоть до 15 апреля 2020 года. Показано, что пик эпидемии еще не достигнут и ожидается в 20-х числах июня (при условии сохранения режима самоизоляции и отсутствия дополнительных ограничительных мер). Для указанных промежутков измерений математическая модель SEIR-D наиболее точно моделирует ситуацию распространения коронавирусной инфекции в Новосибирской области.

Для моделирования и прогнозирования эпидемии COVID-19 не было использовано дополнительных ограничений на параметры моделей, помимо параметра смертности (ограничения на него были проанализированы по сценариям 10 разных стран). Кроме того, не потребовалось кластерных вычислений для получения результатов в рамках двух моделей.


\end{document}